\documentclass[aps,pra,reprint,groupedaddress,superscriptaddress]{revtex4-1}
\usepackage[dvipsnames]{xcolor}
\usepackage{graphicx,amsmath,amssymb}
\usepackage{newtxtext,newtxmath} 
\usepackage{graphicx}
\usepackage[colorlinks,linkcolor=Blue,citecolor=Blue,urlcolor=Blue,breaklinks=true]{hyperref}
\begin{document}
\title{Quantum simulation of discrete curved spacetime by the Bose-Hubbard model: from analog acoustic black hole to quantum phase transition}
\author{F. Bemani}  \email{foroudbemani@gmail.com}
\address{Department of Physics, Faculty of Science, University of Isfahan, Hezar Jerib, 81746-73441, Isfahan, Iran}
\author{R. Roknizadeh} \email{r.roknizadeh@gmail.com}
\author{M. H. Naderi} \email{mhnaderi@phys.ui.ac.ir}
\address{Department of Physics, Faculty of Science, University of Isfahan, Hezar Jerib, 81746-73441, Isfahan, Iran}
\address{Quantum Optics Group, Department of Physics, Faculty of Science, University of Isfahan, Hezar Jerib, 81746-73441, Isfahan, Iran}
\date{\today}

\begin{abstract}
	We present a theoretical scheme to simulate quantum field theory in a discrete curved spacetime based on the Bose-Hubbard model describing a Bose-Einstein condensate trapped inside an optical lattice. Using the Bose-Hubbard Hamiltonian, we first introduce a hydrodynamic presentation of the system evolution in discrete space. We then show that the phase (density) fluctuations of the trapped bosons inside an optical lattice in the superfluid (Mott insulator) state obey the Klein-Gordon equation for a massless scalar field propagating in a discrete curved spacetime. We derive the effective metrics associated with the superfluid and Mott-insulator phases and, in particular, we find that in the superfluid phase the metric exhibits a singularity which can be considered as a the manifestation of an analog acoustic black hole. The proposed approach is found to provide a suitable platform for quantum simulation of various spacetime metrics through adjusting the system parameters. 
\end{abstract}

\maketitle

\section{Introduction}
Two celebrated theories of the modern physics are the general theory of relativity (GTR) and the quantum field theory (QFT). GTR unifies the special theory of relativity and the gravity and it evolves our understanding of the universe around us by providing a geometric interpretation of gravitation. It has some magnificent predictions such as light deflection by gravity, gravitational waves, and black holes \cite{Walecka}. On the other hand, QFT which combines the elements of quantum mechanics and the special theory of relativity, theoretically describes the interactions of the fundamental forces and subatomic particles. QFT in the curved spacetime leads to fascinating predictions such as particle production by the time-dependent gravitational fields, or by the time-independent gravitational fields that contain horizons \cite{Birrell}. QFT in the curved spacetime is regarded as the first approximation to the quantum gravity.

Even for the small values of spacetime curvature, experimental observation of the predictions of  QFT in the curved spacetime seems to be impossible.  To examine such predictions, one can employ the notion of quantum simulation (see \cite{Georgescu} and references therein). Various analog models have been proposed as the simulators for these theories \cite{Novello,Barcelo2005,Unruh,Faccio,Bemani}.  These analog models provide accessible experimental models for quantum field theory in curved space-time. Some properties (mainly kinematic) of GTR can be investigated by the analogies with accessible physical systems  \cite{Novello,Barcelo2005,Unruh,Faccio}.  

The impressive experimental progress in quantum optics, in particular, manipulating and controlling atomic ensembles and the electromagnetic field have opened a route towards the accumulation of the boson particles into the lowest-energy single-particle state, i.e, the formation of BEC \cite{Dalfovo}. Since their first experimental realization in alkalis \cite{Anderson,Davis}, Bose-Einstein condensates (BECs) have been investigated extensively from both theoretical and experimental points of view \cite{Dalfovo}. Several aspects of the GTR and QFT have been analyzed by analogy with a BEC \cite{Garay,Garay2,Barcelo,Mayoral,PAnderson,Finazzi,Girelli,Fedichev1,Fedichev2,Fedichev3,Fischer2004}. The phononic perturbations in a BEC satisfy the Klein-Gordon equation in a curved spacetime and the corresponding metric possesses a singularity which can be regarded as an acoustic analog black hole. Therefore, BECs offer a unique opportunity to explore the basic principles of QFT in the curved spacetime.

By superimposing two or three orthogonal propagating laser beams a spatially periodic pattern is formed. The resulting periodic potential can be used to confine neutral atoms via the Stark shift. Loading ultracold atomic gasses into optical lattices has opened a promising route oriented towards the investigation of strongly correlated quantum many-body systems \cite{Morsch,Jaksch,Lewenstein,Bloch}. There is a close analogy between ultracold atomic gasses inside light-induced periodic potentials and electrons in crystal lattices. The nonlinear atom-atom interactions are responsible for the collective fluid-like behavior in BEC \cite{Morsch,Dalfovo}.  Understanding quantum phase transitions and their underlying physics is one of the goals of many-body physics.  Superfluid to Mott insulator phase transition, first experimentally reported in Ref. \cite{Greiner}, is the most famous example of such phase transitions at zero temperature. The high flexibility and tunability of optical lattices and the condensate provide us a unique opportunity to explore the quantum phase transition \cite{Greiner}.  Long decoherence time (of the order of seconds) makes the trapped bosonic atoms in an optical lattice to be a good candidate for the applications of quantum simulations \cite{Jaksch}. The dynamics of ultracold bosonic atoms in an optical lattice is effectively described by a discrete Bose-Hubbard Hamiltonian \cite{Morsch,Jaksch,Jaksch2,Zwerger}, which describes the tunneling of trapped atoms between the lowest vibrational states of the optical lattice.

Discrete Hamiltonians mainly arising in solid state physics have a great ability to be employed as quantum simulators to investigate QFT on lattice \cite{Giuliani,Katsnelson,Szpak2011,Szpak2012,Szpak2014}. Regardless of the elegance of treating the problems in a discrete spacetime, it has a number of advantages; i) the continuum spacetime can be viewed as the limit of discrete spacetime as the lattice spacing tends to zero; ii) the direct connection to the statistical physics is more obvious. iii) renormalization, scaling, universality, and the role of topology is more apparent in the discrete spacetime; iv) in numerical simulations on a computer, we always approximate the evolution of a continuous field by its finite difference in a discrete space \cite{smit2002introduction}. In this regard, dealing with a discrete geometry may be more beneficial. Electrons in a sheet of graphene have a dynamics analogous to the quantum field propagating in a discrete space \cite{Szpak2014}. The effect of curvature on the current flow lines in elastically deformed graphene sheets has been studied in \cite{Stegmann}. The Dirac Hamiltonian describing electrons and positrons moving in an external field has been simulated by ultracold atoms in bichromatic optical lattices \cite{Szpak2012}.

Inspired by the existing studies on the analogy between the Bose-Hubbard model and a trapped BEC in an optical lattice, and also motivated by the analogy between curved spacetimes and BECs, in the present  contribution, by applying the discrete differential geometry,  we investigate the analog spacetime corresponding to the phase/density fluctuations of a trapped BEC in an optical lattice based on the Bose-Hubbard model. The discrete geometric formulation of the quantum fields is of a great significance to understand the quantum phase transitions occurring in the system. To explore the dynamics of the system, by applying the mean-field approximation, we first derive the discrete equations of motion for the mean field plus fluctuations and then, by using the density-phase representation, we obtain the dynamics of the density as well as the phase fluctuations. In superfluid phase within the hydrodynamic limit, the phase fluctuations  of the quantum field are found to obey the covariant Klein-Gordon equation for a massless scalar field propagating in a curved discrete spacetime with a specified metric. From the point of view of quantum phase transition, the two different phases occurring in the system, i.e., the superfluid and the Mott-insulator phases can be distinguished by their corresponding effective metrics for phase and density fluctuations, respectively. In addition, various effective metrics can be simulated by adjusting the system parameters. 

\section{ Dynamics of the system}\label{Sec:Section2}
We consider an interacting ultracold gas of bosonic atoms trapped in a simple cubic lattice optical lattice. The system is effectively described by the Bose-Hubbard Hamiltonian \cite{Zwerger} 
\begin{eqnarray}
&&\hat {\cal H} =   - \sum\limits_{\left\langle {lmn;l'n'm'} \right\rangle } J\hat b_{lmn}^\dag  {{{\hat b}_{l'n'm'}}}  + \frac{U}{2}\sum\limits_{lmn} {{{\hat n}_{lmn}}} \left( {{{\hat n}_{lmn}} - 1} \right) - \mu \sum\limits_{lmn} {{{\hat n}_{lmn}}}  .
\label{BoseHubbardHamiltonian}
\end{eqnarray}
Here $\left\langle {lmn;l'n'm'} \right\rangle$ stands for the nearest-neighbor sites,  ${\hat {b}}_{nlm}^{\dagger }$ and ${\hat {b}}_{nlm}$ are, respectively, the creation and annihilation operators for a bosonic atom at lattice site $(n,l,m) $ obeying the commutation relation $ [{\hat {b}}_{nlm},{\hat {b}}_{n'l'm'}^{\dagger }]=\delta_{n,n'}\delta_{m,m'}\delta_{l,l'} $, and ${\hat {n}}_{nlm}={\hat {b}}_{nlm}^{\dagger }{\hat {b}}_{nlm}$ is the number operator for site $(n,l,m)$. Moreover, $J$, $\mu$ and $U$ are the tunneling rate between adjacent sites, the chemical potential, and the strength of the on-site interaction, respectively given by 
\begin{align}
J &=  - \int {{d^3}{\bf{r}}{w^*}({\bf{r}} \!-\! {{\bf{r}}_{nlm}})\left[ { -\frac{1}{2M} {\nabla ^2}\! +\! {V_{{\rm{lat}}}}({\bf{r}})} \right]w({\bf{r}} \!-\! {{\bf{r}}_{kmn}})} \, ,
\label{Eq:TunnlingRate} \\
\mu  &=  - \int {{d^3}{\bf{r}}{V_{\rm{ext}}}({\bf{r}})\left| {w({\bf{r}} - {{\bf{r}}_{nlm}})} \right|} ^2\, ,
\label{Eq:ChemicalPotential} \\
U &= G\int {{d^3}{\bf{r}}{{\left| {w({\bf{r}} - {{\bf{r}}_{nlm}})} \right|}^4}} \, ,
\label{Eq:OnSiteInteractionRate} 
\end{align}
where $w({\bf{r}} - {{\bf{r}}_{nlm}})$ are the Wannier functions in the lowest Bloch band. Moreover, $ V_{\rm{lat}} (\bf{r})$ and  $ V_{\rm{ext}} (\bf{r})$ are the lattice periodic potential and the external trapping potential, respectively. $G=4\pi a_s/M $ is the strength of the interactions between different atoms in the BEC determined by the $s$-wave scattering length $a_s$ and the atomic mass $M$. The attractive (repulsive) on-site interaction corresponds to $U<0$ ( $U>0$). In what follows we limit ourselves to considering repulsive atom-atom interaction. Although, here we consider the Bose-Hubbard Hamiltonian for trapped atoms in an optical lattice, but we should notice that the Bose-Hubbard model can be employed to describe some other physical systems such as the systems of strongly interacting photons (for recent reviews see e.g., \cite{Hartmann,Noh}) and coupled optomechanical systems \cite{Huang}. 

The Heisenberg equation of motion for the field operator reads
\begin{multline}
{\partial _t}{{\hat b}_{lmn}}\! =\! iJ({{\hat b}_{l - 1,mn}} \!+\! {{\hat b}_{l + 1,mn}} \!+\! {{\hat b}_{lm - 1,n}} + {{\hat b}_{lm + 1,n}} \!+ \!{{\hat b}_{lm,n - 1}} + {{\hat b}_{lmn + 1}}) \\
- i\frac{U}{2}({{\hat b}_{lmn}}{{\hat n}_{lmn}} + {{\hat n}_{lmn}}{{\hat b}_{lmn}} - {{\hat b}_{lmn}}) + i\mu {{\hat b}_{lmn}} .
\label{Eq:HeisenbergEquation}
\end{multline}
The key concept for the derivation of the curved spacetime on lattice is the central finite differences defined by
\begin{align}
{f_{l \pm 1,m,n}}& = {f_{l,m,n}} \pm a{\Delta _x}{f_{l,m,n}} + \frac{{{a^2}}}{2}\Delta _x^2{f_{l,m,n}} +  \ldots \label{Eq:FiniteDifference1}\\
{f_{l,m \pm 1,n}} &= {f_{l,m,n}} \pm a{\Delta _y}{f_{l,m,n}} + \frac{{{a^2}}}{2}\Delta _y^2{f_{l,m,n}} +  \ldots \label{Eq:FiniteDifference2}\\
{f_{l,m,n \pm 1}} &= {f_{l,m,n}} \pm a{\Delta _z}{f_{l,m,n}} + \frac{{{a^2}}}{2}\Delta _z^2{f_{l,m,n}} +  \ldots 
\label{Eq:FiniteDifference3}
\end{align}
where $ a $ is the lattice spacing. Using definitions (\ref{Eq:FiniteDifference1}--\ref{Eq:FiniteDifference3}) and (7) and neglecting terms
containing higher orders of $ \mathcal{O}(a^2) $, we can write
\begin{align}
\Delta _x^2{f_{lmn}} &= \frac{1}{a^2}({f_{l + 1,m,n}} + {f_{l - 1,mn}} - 2{f_{lmn}}),\label{Eq:FiniteDifference4}\\
\Delta _y^2{f_{lmn}} &= \frac{1}{a^2}({f_{l,m + 1,n}} + {f_{l,m - 1,n}} - 2{f_{lmn}}),\label{Eq:FiniteDifference5}\\
\Delta _z^2{f_{lmn}} &= \frac{1}{a^2}({f_{l,m,n + 1}} + {f_{l,m,n - 1}} - 2{f_{lmn}}),\label{Eq:FiniteDifference6}
\end{align}
Using definitions (\ref{Eq:FiniteDifference4}--\ref{Eq:FiniteDifference6}), we can write Eq.~(\ref{Eq:HeisenbergEquation}) as
\begin{equation}
\partial _t {\hat b}_j = iJ(a^2\Delta ^2{\hat b}_j + 6{\hat b}_j) - i\frac{U}{2} ( {\hat b}_j {\hat n}_j + {\hat n}_j{\hat b}_j - {\hat b}_j ) + i\mu {\hat b}_j\, ,
\label{Eq:DGPE}
\end{equation}
where we have defined ${\Delta ^2} \equiv \Delta _x^2 + \Delta _y^2 + \Delta _z^2 $. Hereafter, the indexes $(l,m,n)$ denoting the spatial dependence of the field operator is replaced by $ j $ for conciseness of notation. 
Eq.~(\ref{Eq:DGPE}) represents a discrete version of the Gross-Pitaevskii equation with an effective mass $m_{\rm{eff}}=(2Ja^2)^{-1}$ describing the trapped atoms in the optical lattice. The effective mass can be controlled by manipulating the lattice depth.  In contrast to the small value of $ J $ which leads to the large mass, a large value of $ J $ provides the atoms a small mass so that they can move freely. 

Within the framework of the mean-field approximation the operator ${\hat b}_j$ can be decomposed into its mean value, $ b_j=\langle {\hat b}_j  \rangle $, and a fluctuation component, $ \delta {{\hat b}_j} $, with zero mean value
\begin{equation}
{\hat b_j} = {b_j} + {\hat \chi _j} = {b_j}(1 + \delta {\hat b_j}){\mkern 1mu} ,
\label{Eq:mean-field}
\end{equation}
where, for subsequent mathematical convenience, we have set $ \hat \chi_j= b_j \delta {\hat b}_j $ with $  \langle \delta {{\hat b}_j}\rangle =0 $.	At zero temperature the fluctuations are essentially due to the quantum noise and at finite temperature, the thermal noise may also be relevant. By inserting Eq.~(\ref{Eq:mean-field}) in Eq.~(\ref{Eq:DGPE}), one may obtain the equations of motion for the mean field and the fluctuation operator which satisfy, respectively
\begin{equation}
\partial _t b_j = iJ(a^2\Delta ^2b_j + 6 b_j) - i\frac{Ub_j}{2} (  2n_j -1 ) + i\mu b_j\, ,
\label{Eq:background}
\end{equation}
\begin{equation}
\partial _t \delta {\hat b}_j \!=\! iJa^2\!\left[\Delta ^2\delta {\hat b}_j \!+\! 2\frac{\Delta b_j}{b_j}\Delta \delta {\hat b}_j \right]\!-\! iUn_j( \delta {\hat b}_j \!+\! \delta \hat b_j^\dag ) \, .
\label{Eq:fluctuations}
\end{equation}
Equation (\ref{Eq:fluctuations}) is analogous to the Bogoliubov-de Gennes equation describing the dynamics of fluctuations in BEC \cite{Dalfovo}.
\begin{figure}
	\begin{center}
		\includegraphics[width=8.5cm]{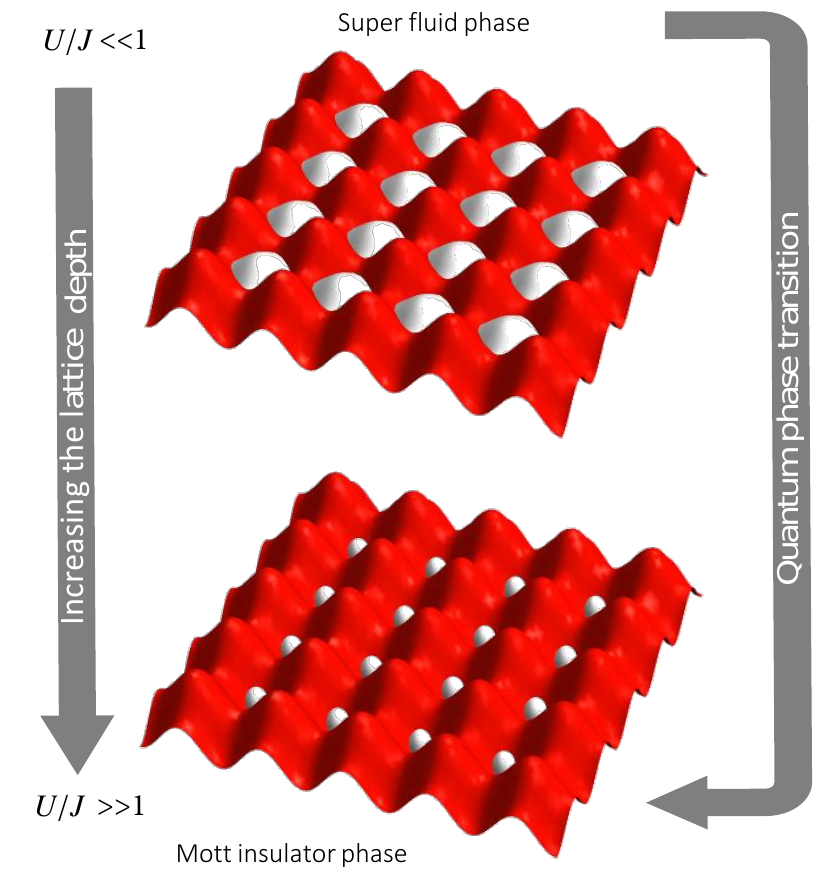}
	\end{center}
	\caption{(Color online). Schematic illustration of  the atoms over optical lattice sites: (top) superfluid and (bottom) Mott insulator states. The phase transition from superfluid to Mott insulator is achieved by increasing the lattice depth.}
	\label{Fig:Fig1}
\end{figure}

\section{Density-phase representation and the effective metric for the fluctuations}
In order to get a clear physical insight into Eqs. (\ref{Eq:background}) and (\ref{Eq:fluctuations}) we express the atomic operator $ {\hat b}_j  $ in the density-phase representation, namely 
\begin{equation}\label{Eq:decompositon}
{\hat b}_j = \sqrt {{\hat n}_j} \exp (i{\hat \phi }_j)
\end{equation} where ${\hat n}_j$ and ${\hat \phi }_j$ denote the amplitude and the phase of the atomic operator, respectively. We should notice that, some issues arise due to the controversial nature of the quantum phase operator. In principle, one could write the quantum field operator as Eq~(\ref{Eq:decompositon}), as long as $n_j=\langle\hat{n}_j\rangle\gg1$. In the limit of large $n_j$, the local number fluctuation and phase operators are conjugate variables and the Bose-Hubbard Hamiltonian is analogous to the quantum rotor model \cite{Javanainen}. Using a self-consistent mean-field expansion for large filling factor, one can investigate the time evolution of quantum fluctuations in the Bose-Hubbard model \cite{Fischer}. In other words the present model will be valid at large filling factor.  One can easily relate the usual decomposition of the atomic operator in linearized regime  to the amplitude and phase fluctuations denoted by  $\delta{\hat n}_j$ and $\delta{\hat \phi }_j$, respectively
\begin{equation}
b_j = \sqrt { n_j} \exp (i \phi _j), \qquad \delta {{\hat b}_j} =  i\delta {\hat \phi }_j + \delta {\hat n}_j/2n_j .
\end{equation}
Substituting these two expressions into Eqs. (\ref{Eq:background}) and (\ref{Eq:fluctuations}) leads to the equations of motion for the mean density and the mean phase 
\begin{equation}
{\partial _t}{n_j} + \Delta .\left( {{n_j}{\bf{v}}_j} \right) = 0\, , \label{Eq:hydrodynamics1}
\end{equation}
\begin{equation}
\partial _t \phi _j =  - J{a^2}\left[ (\Delta \phi _j)^2 - \frac{\Delta ^2n_j}{2n_j} + \frac{{{{(\Delta {n_j})}^2}}}{{4n_j^2}} \right] 
 + (6J + U/2 + \mu ) -U{n_j},
\label{Eq:hydrodynamics2}
\end{equation}
where we have defined ${{\bf v}_j}=2Ja^2 \Delta \phi_j$ to be the local velocity of the fluid. Equation (\ref{Eq:hydrodynamics2}) can be transformed into the following equation
\begin{equation}
{\partial _t}{{\bf v}_j} + {{\bf v}_j}.\Delta {{\bf v}_j} =   Ja^2\Delta (\mu  +6J+U/2+ V_{\rm{Q}}- U{n_j})\, ,
\label{Eq:hydrodynamics3}
\end{equation}
where $V_{\rm{Q}} = {J^2}({\Delta ^2}\sqrt {{n_j}} )/\sqrt {n_j} $ is the so-called ``quantum potential''. Equations (\ref{Eq:hydrodynamics1}) and (\ref{Eq:hydrodynamics3}) are the mass and momentum conservation equations for the trapped bosons in optical lattice, respectively. The equations of motion for the density fluctuation and the phase fluctuation are obtained as
\begin{eqnarray}
{\partial _t}\delta {{\hat n}_j} =  - \Delta ({{\bf{v}}_j}\delta {{\hat n}_j} + 2J{a^2}{n_j}\Delta \delta {{\hat \phi }_j}), \label{Eq:hydrodynamics4}
\end{eqnarray}
\begin{eqnarray}
\partial _t\delta {\hat \phi }_j =  - {\bf{v}}_j\Delta \delta {\hat \phi }_j - \frac{{c_j^2}}{{2J{a^2}n_j}}\delta {{\hat n}_j} +\frac{{c_j^2}}{{8J{a^2}{n_j}}}{\xi_j ^2}\Delta [{n_j}\Delta (\frac{{\delta {{\hat n}_j}}}{{{n_j}}})], \label{Eq:hydrodynamics5}
\end{eqnarray}
where $c_j = a \sqrt {2Jn_jU}$ and $\xi_j  = a \sqrt{ 2J/(n_j U)} $ are the local speed of excitations and the so-called healing length, respectively. 
	In the limit of large filling factor $n_j$, one can easily generalize the results of \cite{Javanainen,Fischer} to a 3D optical lattice. Therefore, the dispersion relation for waves with frequency $ \omega({\bf k}) $ and wavevector $ {\bf k}= k_x\hat{i}+k_y\hat{j}+k_z\hat{k} $, is given by
\begin{equation}
\omega \left( {{k_x},{k_y},{k_z}} \right) = \sum\limits_{l = x,y,z} {4n_jUJ{{\sin }^2}\left( {\frac{{{k_l}a}}{2}} \right) + 4{J^2}{{\sin }^4}\left( {\frac{{{k_l}a}}{2}} \right)}. 
\label{Eq1}
\end{equation}  
We notice that for $ |{\bf k}| a <<1 $ it reduces to the well-known Bogoliubov spectrum 
\begin{equation}
\omega \left( {{k_i},{k_j},{k_k}} \right) = \sum\limits_{l = i,j,k} {\left( {n_jUJ{a^2} + \frac{{{J^2}{a^4}}}{4}k_l^2} \right)} k_l^2.
\label{Eq2}
\end{equation}
In this paper, our goal is to simulate a continuum spacetime via a discrete spacetime and therefore we focus our attention on the case of long wavelength fluctuations (hydrodynamic limit). Considering the Bogoliubov mean-field approximation for the fluctuations of the weakly depleted condensate the excitation spectrum given by Eq~(\ref{Eq1}) becomes gapless.
\subsection{Superfluid phase}
\begin{figure}
	\begin{center}
		\includegraphics[width=8.5cm]{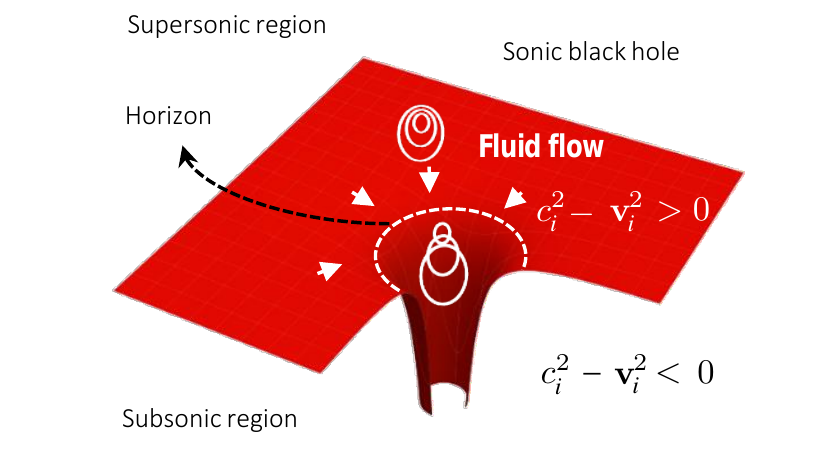}
	\end{center}
	\caption{(Color online). The generation of an acoustic black hole in the analog spacetime in the superfluid state corresponding to the metric of Eq.~(\ref{Eq:Metric}).}
	\label{Fig:Fig2}
\end{figure}
It is well-known that the nature of many-body eigenstates of the Bose-Hubbard Hamiltonian in Eq.~(\ref{BoseHubbardHamiltonian}) is dependent on the ratio of the on-site interaction energy $ U $ to the tunneling energy  $ J $. An experimental way for controlling this ratio is based on the Feshbach resonances method \cite{Chin} in which the scattering length is controlled by an external magnetic field. Alternatively, the dimensionless parameter $U/J$  can be controlled by manipulating the depth of the optical lattice \cite{Islam}. The competition between minimizing the tunneling energy and the on-site atom-atom interaction manifests itself as the well-known superfluid-Mott insulator phase transition in the ground state of the Bose-Hubbard model (Fig.~(\ref{Fig:Fig1})). In the regime $U/J\ll1$ which corresponds to the superfluid phase, atoms tend to accumulate in a single quantum state and delocalize over the entire lattice sites. The many-particle state of the superfluid can be written as
\begin{equation}
{\left| \Psi  \right\rangle _{\rm{SF}}} \propto \prod\limits_j {\left| {{\alpha _j}} \right\rangle } \, ,
\end{equation}
i.e., the state vector at each site can be described as the Glauber coherent state  $ {\left| {{\alpha _j}} \right\rangle } $. 
In the superfluid phase, in the hydrodynamic approximation, i.e., in the regime where the characteristic length of the spatial variations of the condensate density is  much larger than $\xi_j$ the last term in Eq.~(\ref{Eq:hydrodynamics5}) can be safely ignored and thus we have 
\begin{equation}\label{Eq:hydrodynamics60}\delta {{\hat n}_j} \simeq  - 2J a^2n_j\left( {{\partial _t}\delta {{\hat \phi }_j} + {{\bf{v}}_j}\Delta \delta {{\hat \phi }_j}} \right)/c_j^2\, .
\end{equation}
Combining Eqs. (\ref{Eq:hydrodynamics4}) and (\ref{Eq:hydrodynamics60}) results in the following equation for the phase fluctuations of the quantum field
\begin{equation}\label{eqfase}
- ({\partial _t} + \Delta {{\bf{v}}_j})\frac{{J{a^2}{n_j}}}{{c_j^2}}({\partial _t} + {{\bf{v}}_j}\Delta )\delta {{\hat \phi }_j} + \Delta J{a^2}{n_j}\Delta \delta {{\hat \phi }_j} = 0\;.
\end{equation}
These fluctuations, within the hydrodynamic approximation, are analogous to the collective quantum field on a curved metric. Therefore, the phase fluctuations of the quantum field obey the covariant Klein-Gordon equation for a massless scalar field propagating in a curved discrete spacetime
\begin{equation}
\Box \delta \hat \phi_j = \frac{1}{\sqrt{-g}}\Delta_\mu (\sqrt{-g}g^{\mu\nu}\Delta_\nu \delta \hat \phi_j )=0\;,
\label{box}
\end{equation}
where $g^{\mu \nu}$ is the inverse curved spacetime metric and $g$ is the determinant of the metric $g_{\mu \nu}$
\begin{equation}
g_{\mu \nu }^{SF} = \sqrt {2{n_j}J{a^2}/U}  \left[ {\begin{array}{*{20}{c}}
	{ - \left( {c_j^2 - {{\bf{v}}_j}.{{\bf{v}}_j}} \right)}&{ - v_j^l}\\
	{ - v_j^m}&{{\delta _{lm}}}
	\end{array}} \right].
\label{Eq:Metric}
\end{equation}
Here, the Greek summation indices range from $0$ to $2$ and the indices $(l,m)$ range from $1$ to $2$.
The line element of this spacetime is given by
\begin{align}
\Delta {s^2} &= {g_{\mu \nu ,j}}\Delta {x^\mu }\Delta {x^\nu } \nonumber\\
& = \!\sqrt{\frac{{2{n_j}J {a^2} }}{U}}\! \left[ {\! - {c^2}d{t^2} \!+\! (\Delta {{\bf{x}}_j} \!-\! {{\bf{v}}_j}dt)(\Delta {{\bf{x}}_j} \!-\! {\bf{v}}_jdt)} \right].
\end{align}
At the point $c_j=|\bf{v}_j|$ or, equivalently, when $ U/J = 2{\left( {{\phi _{i + 1}} - {\phi _j}} \right)^2}/{n_j} $ the metric of Eq.~(\ref{Eq:Metric}) exhibits a singularity. The meaning of singularities in fluid dynamics and general relativity has been discussed in \cite{Cadoni}. We refer this condition as an acoustic analog black hole (depicted in Fig.~(\ref{Fig:Fig2})). In such a situation, the sound waves traveling with $c_j<|\bf{v}_j|$ are trapped inside the \textit{supersonic} region and they are not able to propagate backward into the \textit{subsonic} region. Since both sound and the fluid velocity can be controlled by the lattice depth, one can generate an analog black hole in the system. The generated analog black hole is the turning point to the superfluid  Landau instability \cite{Raman}. For a homogeneous lattice $ \Delta \phi_j=0 $ or equivalently $ \textbf{v}_j=0 $ the effective metric has the form
\begin{equation}
g_{\mu \nu }^{H} = \sqrt {2 n_j J a^2/U}  \left[ {\begin{array}{*{20}{c}}
	{ - c_j^2} &{0}\\
	{ 0}&{{\delta _{lm}}}
	\end{array}} \right],
\label{Eq:MetricHomogeneous}
\end{equation}
which is suitable for the simulation of various analog metrics in the curved spacetime.

\subsection{Mott insulator phase}
Starting from the superfluid phase (see Fig.~(\ref{Fig:Fig1})), one can reach the Mott insulator phase by increasing the lattice depth. The Mott insulator regime emerges when the tunneling energy is much smaller than the on-site repulsion energy,  i.e., $U/J\gg1$. In this regime, all atoms are completely localized over the lattice \cite{Greiner}.
The many-body state vector of the ground state can be expressed as 
\begin{equation}
{\left| \Psi  \right\rangle _{\rm{MI}}} \propto \prod\limits_j {\left| {{n_j}} \right\rangle } \, ,
\end{equation}
where ${\left| {{n_j}} \right\rangle }$ is the eigenstate of the number operator of particles on the $j$th site,  $\hat{n}_j$. As stated in the beginning of the section, we consider large filling factor $n_j\gg1$ for the Mott insulator regime.
Rewriting Eq.~(\ref{Eq:hydrodynamics5}) in terms of $ J $ and $ U $ leads to
\begin{equation}
{\partial _t}\delta {{\hat \phi }_j} = J{a^2}\left[ { - 2\Delta {\phi _j}\Delta \delta {{\hat \phi }_j} + \frac{1}{{2{n_j}}}\Delta [{n_j}\Delta (\frac{{\delta {{\hat n}_j}}}{{{n_j}}})]} \right] - U\delta {{\hat n}_j},
\label{Eq:MottHydrodynamics}
\end{equation}
In the Mott insulator regime, $U\gg J$, one can neglect the first term in Eq~(\ref{Eq:MottHydrodynamics}), approximately and thus
\begin{equation}
{\partial _t}\delta {{\hat \phi }_j} = - U\delta {{\hat n}_j}, \label{Eq:hydrodynamics6}
\end{equation}
By performing the time derivative on Eq.~(\ref{Eq:hydrodynamics4}), and using Eqs.~(\ref{Eq:hydrodynamics1}), (\ref{Eq:hydrodynamics2}) and (\ref{Eq:hydrodynamics6}) we readily get
\begin{eqnarray}\label{Eq:hydrodynamics7}
\partial _t^2\delta {{\hat n}_j} =  - 2J{a^2}\Delta (\delta {{\hat n}_j}\Delta {\partial _t}{\phi _j} + \Delta {\phi _j}{\partial _t}\delta {{\hat n}_j} + \Delta \delta {{\hat \phi }_j}{\partial _t}{n_j}  - U{n_j}\Delta \delta {{\hat n}_j}),
\end{eqnarray}
Assuming $\Delta {n_j} \ll {n_j} $, and keeping only the last term in the parentheses one has
\begin{equation}\label{Eq:hydrodynamics7}
\partial _t^2\delta {{\hat n}_j} - \Delta (c_j^2 \Delta \delta {{\hat n}_j}) = 0,
\end{equation}  

Thus, in this regime, the effective spacetime for the density fluctuations is given by the metric $ g_{\mu \nu }^{MI} =g_{\mu \nu }^{H} $
\subsection{Realization of various metrics in the Bose-Hubbard model}
Finally, we consider three other examples of the realization of spacetime metrics. The fist example is the Friedmann-Lema\^{\i}tre-Robertson-Walker (FLRW) metric \cite{Walecka,Birrell} which is an exact solution of Einstein's field equations of GTR which describes a homogeneous and isotropic universe on a surface of constant time, but it is no longer static (time-dependent metric). The FLRW metric corresponds to a spacetime geometry defined by the line element
\begin{equation}\label{Eq:FLRW}
ds^2_{\rm{FLRW}}=-c^2dt^2+R^2(t) \left[ dx^2 +dy^2+dz^2\right]\, ,
\end{equation}
where $c$ is the speed of light and $ R(t) $ is the scale factor. With properly modulating the tunneling energy $ J(t)=J_0\exp(-Ht)$, one can simulate FLRW metric with $ R(t)= \exp(Ht/2) $ in the superfluid phase. A modulated tunneling energy leads to a time-dependent effective mass $ m\left( t \right) = {(2J_0a^2)^{-1}}{\exp{(Ht)}}$. According to Eq.~(\ref{Eq:TunnlingRate}), one can modulate the tunneling energy by modulating the optical lattice (exponential modulation of the tunneling energy was discussed in \cite{Fischer,Schutzhold}). The dependence of the effective mass on time and space can be studied by introducing the mass tensor (the details of the analysis is given in ref. \cite{Barcelo}).  In the present case, the effective mass depends only on time and hence the metric of Eq.~(\ref{Eq:Metric}) is generalized as
\begin{equation}
g_{\mu \nu }^{SF} = \sqrt {2{n_j}{J_0}{a^2}/U} \left[ {\begin{array}{*{20}{c}}
	{ - \left( {c_j^2 - {{\bf{v}}_j}.{{\bf{v}}_j}} \right)}&{ - v_j^l}\\
	{ - v_j^m}&{{e^{Ht}}{\delta _{lm}}}
	\end{array}} \right].
\end{equation} 
In the superfluid phase, for a homogeneous lattice the line element can be written as
\begin{equation}
\Delta {s^2} = \sqrt {\frac{{2{n_j}{J_0}{a^2}}}{U}} \left[ { - c_j^2d{t^2} + {e^{Ht}}\left( {\Delta {x^2} + \Delta {y^2} + \Delta {z^2}} \right){\mkern 1mu} } \right].
\end{equation}
As another example let us consider a  homogeneous superfluid. The metric of the corresponding  spacetime is a Minkowski metric given by
\begin{equation}
{\eta _{\mu \nu }} = \sqrt {2nJ{a^2}/U} \left[ {\begin{array}{*{20}{c}}
	{ - {c^2}}&0\\
	0&{{\delta _{lm}}}
	\end{array}} \right]{\mkern 1mu} \, .
\end{equation}
In general the metric ${g_{\mu \nu }}$ can be decomposed into a flat metric given by $ \eta_{\mu \nu} $ and a curved spacetime metric given by $ {h_{\mu \nu }}$. A small deviation from homogeneity in the background field, $ {n_j} = n + {\varepsilon _j}$, manifests itself as a curved spacetime in the metric  $ {h_{\mu \nu }} $ given by   
\begin{equation}
{h_{\mu \nu }} = \sqrt {{Ja^2}/{{(2nU)}}} {\varepsilon _j}\left[ {\begin{array}{*{20}{c}}
	{ - 3{c^2}}&0\\
	0&{{\delta _{lm}}}
	\end{array}} \right]\, .
\end{equation}
Therefore, one can obtain analog curved spacetime in the superfluid phase by manipulating the background density. Again one can also use the Mott insulator phase metric $g_{\mu \nu }^{MI}  $ to mimic the curved space time. In other words, any inhomogeneity in the background can be translated into the spacetime curvature.

As the third example, we discuss the generation and propagation of analog gravitational radiation in the trapped atoms inside an optical lattice. As in the previous example, we set $h_{\mu\nu}$ to be a symmetric tensor field (with respect to the swapping of its indices) defined on a flat Minkowski background spacetime. In this case, the tensor field can vary rapidly with time. A 1D gravitational wave metric corresponds to a spacetime geometry defined by the line element
\begin{equation}\label{Eq:GW}
ds^2_{\rm{GW}}=-c^2dt^2+(1+h_+(t))  dx^2  ,
\end{equation}
where $h_{+} $  is a time-dependent function describing a gravitational wave with $+$ polarization. Choosing a modulated tunneling energy according to $ J(t)=J_0 (1-\epsilon_+\sin\nu_+t) $ where $ \epsilon_+ $ is a small amplitude and $ \nu_+ $ is the frequency of the modulation leads to a time-dependent effective mass $ m(t)\simeq {(2J_0a^2)^{-1}}(1+\epsilon_+\sin\nu_+t) $. Therefore,  the metric of Eq.~(\ref{Eq:Metric}) for a homogeneous 1D lattice with a modulated tunneling energy  $ J(t)=J_0 (1-\epsilon_+\sin\nu_+t) $ can be written as
\begin{equation}
g_{\mu \nu }^{SF} = \sqrt {2{n_j}{J_0}{a^2}/U} \left[ {\begin{array}{*{20}{c}}
	{ -  c_j^2  }&0\\
	0&{{(1+\epsilon_+\sin\nu_+t) }}
	\end{array}} \right].
\end{equation} 
Its corresponding line element can also be written as
\begin{equation}
\Delta {s^2} = \sqrt {\frac{{2{n_j}{J_0}{a^2}}}{U}} \left[ { - c_j^2d{t^2} +   (1+\epsilon_+\sin\nu_+t)\Delta {x^2} } \right].
\end{equation}
which is the same as the line element given by (\ref{Eq:GW}) up to a conformal factor.
\section{Conclusions}
In summary, we have formulated quantum field theory in discrete spacetime based on the Bose-Hubbard Hamiltonian describing a trapped BEC in an optical lattice. The corresponding phase fluctuations of the field operator can be well described by the Klein-Gordon equation for a massless particle propagating in a curved discrete spacetime. The superfluid and the Mott insulator phases are associated with two different metrics.  If trapped atoms in an optical lattice act as a superfluid, they must exhibit the known features of superfluids, including their analogy to space-time geometries. The emergence of the phase transition from the superfluid to Mott insulator depends on the system parameters. In the Mott insulator phase, the phase at each lattice site is random, so it only makes sense to consider amplitude fluctuations. The amplitude fluctuations also satisfy a space-time geometry, but it is quite obvious that this geometry cannot couple space and time, since tunneling is strongly suppressed in the Mott insulator. 
In addition, various analog metrics can be simulated by adjusting the system parameters. The proposed approach provides a suitable platform for quantum simulation of continuous quantum fields in curved spacetimes by their counterparts in discrete spacetimes.  Moreover, the present study can be extended to some other physical systems in which the Bose-Hubbard model is realizable, such as strongly interacting photons \cite{Hartmann,Noh} and coupled optomechanical systems \cite{Huang}.
\section*{Acknowledgments}
We wish to thank the Office of Graduate Studies of
the University of Isfahan for its support. 
%

\end{document}